\documentclass[conference]{IEEEtran}

\usepackage[
  pass,
]{geometry}

\usepackage[latin1]{inputenc}
\usepackage{amsmath,bbm,subfigure}

\usepackage{graphicx}
\usepackage{acronym}
\usepackage{changes}
\usepackage{lipsum}
\definechangesauthor[name={Authors}, color=blue]{authors}

\usepackage{acronym}

\usepackage[acronym]{glossaries}
\usepackage{glossary-mcols}
\usepackage{url,comment}
\usepackage{booktabs}
\usepackage{flushend}

\usepackage{amsmath}
\usepackage{algorithm}
\usepackage[noend]{algpseudocode}
\usepackage{siunitx} 

\usepackage{array}
\newcolumntype{L}[1]{>{\raggedright\let\newline\\\arraybackslash\hspace{0pt}}m{#1}}
\newcolumntype{C}[1]{>{\centering\let\newline\\\arraybackslash\hspace{0pt}}m{#1}}
\newcolumntype{R}[1]{>{\raggedleft\let\newline\\\arraybackslash\hspace{0pt}}m{#1}}

\makeatletter
\def\BState{\State\hskip-\ALG@thistlm}
\makeatother

\newacronym{AAA}{AAA}{Authentication, Authorization, and Accounting}
\newacronym{RSC}{RSC}{Recursive  Systematic Convolutional}
\newacronym{LLR}{LLR}{Log-Likelihood Ratio}
\newacronym{FS}{FS}{Functional Split}
\newacronym{BBU}{BBU}{Base Band Unit}
\newacronym{COTS}{COTS}{Commercial off-the-shelf }
\newacronym{VNF}{VNF}{Virtualized Network Function}
\newacronym{VNF FG}{VNF FG}{VNF Forwarding Graph}
\newacronym{NFV}{NFV}{Network Function Virtualization}
\newacronym{GPP}{GPP}{General Purpose Processor}
\newacronym{vEPC}{vEPC}{virtual Evolved Packet Core}
\newacronym{LTE}{LTE}{Long Term Evolution}
\newacronym{uRLLC}{uRLLC}{Ultra-Reliable Low-Latency Communications}
\newacronym{eMBB}{eMBB}{enhanced Mobile BroadBand}
\newacronym{mMTC}{mMTC}{massive Machine Type Communications}
\newacronym{OSM}{OSM}{Open Source MANO}
\newacronym{C-EPC}{C-EPC}{Cloud-EPC}
\newacronym{EPCaaS}{EPCaaS}{EPC as a Service}
\newacronym{TDD}{TDD}{Time Division Duplex}
\newacronym{UE}{UE}{User Equipment}
\newacronym{HARQ}{HARQ}{Hybrid Automatic Repeat-Request}
\newacronym{PRB}{PRB}{Physical Resource Blocks}
\newacronym{MCS}{MCS}{Modulation and Coding Scheme}
\newacronym{CQI}{CQI}{Channel Quality Indicator}
\newacronym{DC}{DC}{Dedicated Core}
\newacronym{RR}{RR}{Round Robin}
\newacronym{G}{G}{Greedy}
\newacronym{VPN}{VPN}{Virtual Private Network}
\newacronym{MPLS}{MPLS}{Multiprotocol Label Switching}
\newacronym{OWL}{OWL}{Web Ontology Language}
\newacronym{NST}{NST}{Network Slice Template}
\newacronym{NSST}{NSST}{Network Slice Subnet Template}
\newacronym{NSMF}{NSMF}{Network Slice Management Function}
\newacronym{NSSMF}{NSSMF}{Network Slice Subnet Management Function}
\newacronym{CSMF}{CSMF}{Communication Service Management Function}
\newacronym{FCAPS}{FCAPS}{Fault-management Configuration Accounting Performance and Security}
\newacronym{RF}{RF}{Radio Frequency}
\newacronym{RIC}{RIC}{RAN Intelligent Controller}

\newacronym{CNF}{CNF}{Cloud-Native Network Function}
\newacronym{PLMN}{PLMN}{Public Land Mobile Network}
\newacronym{CLAMP}{CLAMP}{Closed Loop Automation Management Platform}

\newacronym{FDD}{FDD}{Frequency Division Duplex}
\newacronym{OFDM}{OFDM}{Orthogonal Frequency Division Multiplexing}

\newacronym{VM}{VM}{Virtual Machine}
\newacronym{PDCP}{PDCP}{Packet Data Convergence Protocol}
\newacronym{MAC}{MAC}{Medium Access Control}
\newacronym{RLC}{RLC}{Radio Link Control}
\newacronym{RRC}{RRC}{Radio Resource Control}
\newacronym{AM}{AM}{Acknowledged Mode}
\newacronym{UM}{UM}{Unacknowledged Mode}
\newacronym{TM}{TM}{Transparent Mode}
\newacronym{MIMO}{MIMO}{Multiple Input Multiple Output}
\newacronym{MISO}{MISO}{Multiple Input Single Output}
\newacronym{SIMO}{SIMO}{Single Input Multiple Output}
\newacronym{SISO}{SISO}{Single Input Single Output}
\newacronym{MCC}{MCC}{Mobile Country Code}
\newacronym{MNC}{MNC}{Mobile Network Code}
\newacronym{S-TMSI}{S-TMSI}{Shortened Temporary Mobile Subscriber Identity}
\newacronym{IMSI}{IMSI}{International Mobile Subscriber Identity}
\newacronym{DRB}{DRB}{Dedicated Radio Bearer}
\newacronym{GUMMEI}{GUMMEI}{Globally Unique MME Identity}

\newacronym{PCI}{PCI}{Physical-layer Cell Identity}
\newacronym{ROHC}{ROHC}{Robust Header Compression}
\newacronym{SN}{SN}{Sequence Number}
\newacronym{RAR}{RAR}{Random Access Response}
\newacronym{C-RNTI}{C-RNTI}{Cell Radio Network Temporary Identifier}
\newacronym{BSR}{BSR}{Buffer Status Report}
\newacronym{DRX}{DRX}{Discontinuous Reception}
\newacronym{PHR}{PHR}{Power Head Room}
\newacronym{PUSCH}{PUSCH}{Physical Uplink Shared Channel}
\newacronym{ADM}{ADM}{Activation/Deactivation MAC}
\newacronym{GP}{GP}{Gap Period}

\newacronym{RE}{RE}{Resource Element}
\newacronym{RB}{RB}{Resource Block}
\newacronym{REG}{REG}{Resource Element Group}
\newacronym{CSRS}{CSRS}{Cell-Specific Reference Signal}
\newacronym{IFFT}{IFFT}{Inverse Fast Fourier Transform}
\newacronym{OFDMA}{OFDMA}{Orthogonal Frequency Division Multimple Access}
\newacronym{CRC}{CRC}{Cyclic Redundancy Check}
\newacronym{SFC}{SFC}{Service Function Chain}

\newacronym{eNB}{eNB}{Evolved NodeB}
\newacronym{RAN}{RAN}{Radio Access Network}
\newacronym{ARQ}{ARQ}{Automatic Repeat reQuest}
\newacronym{NAS}{NAS}{Non-Access Stratum}
\newacronym{MME}{MME}{Mobility Management Entity}
\newacronym{MIB}{MIB}{Master Information Block}
\newacronym{SIB}{SIB}{System Information Block}
\newacronym{RSRP}{RSRP}{Reference Signal Received Power}
\newacronym{RAT}{RAT}{Radio Access Technologie}
\newacronym{ACK}{ACK}{Acknowledge}
\newacronym{NACK}{NACK}{Negative acknowledge}
\newacronym{PDCCH}{PDCCH}{Physical Downlink Control Channel}
\newacronym{SAW}{SAW}{Stop and Wait}
\newacronym{TTI}{TTI}{Transmission Time Interval}
\newacronym{RRH}{RRH}{Radio Remote Head}
\newacronym{SNIR}{SNIR}{Signal-to-Noise-plus-Interference Ratio}
\newacronym{WCET}{WCET}{Worst Case Execution Time}
\newacronym{GPC}{GPC}{General Purpose Computer}
\newacronym{KPI}{KPI}{Key Performance Indicator}
\newacronym{OAI}{OAI}{Open Air Interface}
\newacronym{IMS}{IMS}{IP Multimedia Subsystem}
\newacronym{vIMS}{vIMS}{virtual IP Multimedia Subsystem}
\newacronym{EPC}{EPC}{Evolved Packet Core}
\newacronym{SDN}{SDN}{Software Defined Network}
\newacronym{C-RAN}{C-RAN}{Centralized-RAN}
\newacronym{OS}{OS}{Operating System}
\newacronym{TB}{TB}{Transport Block}
\newacronym{TBS}{TBS}{Transport Block Size}
\newacronym{QCI}{QCI}{QoS Channel Indicator}

\newacronym{GPU}{GPU}{Graphics Processing Unit}
\newacronym{CPU}{CPU}{Central Processing Unit}

\newacronym{SDU}{SDU}{Service Data Unit}
\newacronym{CBS}{CBS}{Code Block Size}
\newacronym{CB}{CB}{Code Block}
\newacronym{SPMD}{SPMD}{Single Program Multiple Data}
\newacronym{SIMD}{SIMD}{Single Instruction Multiple Data} 

\newacronym{SINR}{SINR}{Signal-to Interference Noise Ratio}
\newacronym{CO}{CO}{Central Office}
\newacronym{CA}{CA}{Carrier Aggregation}
\newacronym{SRS}{SRS}{Sound Reference Signal}
\newacronym{SC-OFDMA}{SC-OFDMA}{Single Carrier - Orthogonal Frequency Division Multiple Access}
\newacronym{FPGA}{FPGA}{Field-Programmable Gate Array}
\newacronym{TA}{TA}{Time Advancing}
\newacronym{CoMP}{CoMP}{Coordinated Multi-point}
\newacronym{NPRB}{NPRB}{Number of Physical Resource Blocks}
\newacronym{RTT}{RTT}{Round Trip Time}
\newacronym{CPRI}{CPRI}{Common Public Radio Interface}
\newacronym{CBR}{CBR}{Constant Bit Rate}
\newacronym{NRB}{NRB}{Number of Resource Blocks}
\newacronym{BJF}{BJF}{Biggest Job First}
\newacronym{EDF}{EDF}{Earliest Deadline First}
\newacronym{FCFS}{FCFS}{First-come, First-served}
\newacronym{PSTN}{PSTN}{Public Switched Telephone Network}
\newacronym{ETSI}{ETSI}{European Telecommunications Standards Institute}
\newacronym{vBBU}{vBBU}{virtualized BBU}
\newacronym{vRAN}{vRAN}{virtualized RAN}
\newacronym{IoT}{IoT}{Internet of Things}
\newacronym{B2B}{B2B}{Business to Business}
\newacronym{B2C}{B2C}{Business to Customer}
\newacronym{QoE}{QoE}{Quality of Experience}
\newacronym{QoS}{QoS}{Quality of Service}
\newacronym{VNO}{VNO}{Virtual mobile Network Operator}
\newacronym{SLA}{SLA}{Service Level Agreement}
\newacronym{VRRM}{VRRM}{Virtual Radio Resource Management}
\newacronym{KVM}{KVM}{Kernel-based Virtual Machine}
\newacronym{LXC}{LXC}{Linux Containers}
\newacronym{PS}{PS}{Processor Sharing}

\newacronym{eCPRI}{eCPRI}{evolved CPRI}
\newacronym{RoE}{RoE}{Radio over Ethernet}
\newacronym{PAPR}{PAPR}{Peak-to-average power ratio}
\newacronym{SC-FDMA}{SC-FDMA}{Single Carrier Frequency Division Multiple Access}
\newacronym{AGC}{AGC}{Automatic Gain Control}
\newacronym{PMD}{PMD}{Polarization Mode Dispersion}
\newacronym{ADC}{ADC}{Analogic-Digital Converter}

\newacronym{IQ}{IQ}{In-Phase Quadrature}
\newacronym{xRAN}{xRAN}{extensible Radio Access Network}
\newacronym{ISI}{ISI}{Inter-symbol interference}
\newacronym{FFT}{FFT}{Fast Fourier Transform}
\newacronym{IPC}{IPC}{Inter process communication}
\newacronym{CCDU}{CCDU}{Channel Coding Data Unit}
\newacronym{CC}{CC}{Channel Coding}
\newacronym{gNB}{gNB}{next-Generation Node B}
\newacronym{EUTRAN}{EUTRAN}{Evolved Universal Terrestrial Radio Access Network}
\newacronym{SCTP}{SCTP}{Stream Control Transmission Protocol}
\newacronym{NR}{NR}{New Radio}
\newacronym{NF}{NF}{Network Function}
\newacronym{CU}{CU}{Central Unit}
\newacronym{DU}{DU}{Distributed Unit}
\newacronym{NGC}{NGC}{Next Generation Core}
\newacronym{DL}{DL}{down-link}
\newacronym{UL}{UL}{up-link}
\newacronym{LJF}{LJF}{Largest Job First}
\newacronym{RANaaS}{RANaaS}{RAN as a Service}
\newacronym{NaaS}{NaaS}{Network as a Service}
\newacronym{NS}{NS}{Network Service}
\newacronym{FG}{FG}{Forwarding Graph}
\newacronym{VNFC}{VNFC}{VNF Component}
\newacronym{MANO}{MANO}{Management and Orchestration}
\newacronym{FIFO}{FIFO}{First In Firs Out}
\newacronym{NFVI}{NFVI}{NFV Infrastructure}
\newacronym{NFVO}{NFVO}{NFV Orchestrator}
\newacronym{PoP}{PoP}{Point of Presence}
\newacronym{NAT}{NAT}{Network Address Translation}
\newacronym{CDN}{CDN}{Content Delivery Network}
\newacronym{VNFM}{VNFM}{VNF Manager}
\newacronym{EM}{EM}{Element Management}
\newacronym{VIM}{VIM}{Virtualised Infrastructure Manager}

\newacronym{e2e}{e2e}{end-to-end}

\newacronym{AMF}{AMF}{Access and Mobility Management Function}
\newacronym{SMF}{SMF}{Session Management Function}
\newacronym{UPF}{UPF}{User Plane Function}
\newacronym{PCF}{PCF}{Policy Control Function}
\newacronym{UDM}{UDM}{Unified Data Management}
\newacronym{NRF}{NRF}{NF Repository Function}
\newacronym{AUSF}{AUSF}{Authentication Server Function}
\newacronym{API}{API}{Application Programming Interface}
\newacronym{HSS}{HSS}{Home Subscriber Server}
\newacronym{PCRF}{PCRF}{Policy and Charging Rules Function}
\newacronym{SOA}{SOA}{Software-Oriented Architecture}
\newacronym{AKA}{AKA}{Authentication and Key Agreement}
\newacronym{AF}{AF}{Application Function}
\newacronym{NEF}{NEF}{Network Exposure Function}
\newacronym{NSSF}{NSSF}{Network Slice Selection Function}
\newacronym{NSSP}{NSSP}{Network Slice Service Profile}
\newacronym{VES}{VES}{Virtual Event Streaming}
\newacronym{NSSAI}{NSSAI}{Network Slice Selection Assistance Information}

\newacronym{NSSI}{NSSI}{Network Slice Subnet Instance}

\newacronym{NSS}{NSS}{Network Slice Subnet}
\newacronym{NSC}{NSC}{Network Slice Customer}
\newacronym{NSP}{NSP}{Network Slice Provider}
\newacronym{CSC}{CSC}{Communication Service Customer}
\newacronym{CSP}{CSP}{Communication Service Provider}
\newacronym{SST}{SST}{Slice/Service Type}
\newacronym{SD}{SD}{Slice Differentiator}
\newacronym{USRP}{USRP}{UE Router Selection Policy}
\newacronym{S-NSSAI}{S-NSSAI}{Single Network Slice Selection Assistance Information}
\newacronym{ONISTT}{ONISTT}{Open Net-centric Interoperability Standards for Training and Testing}
\newacronym{KB}{KB}{Knowledge Base}
\newacronym{NSI}{NSI}{Network Slice Instance}

\newacronym{VF}{VF}{Virtual Function}
\newacronym{VFC}{VFC}{Virtual Function Component}
\newacronym{CR}{CR}{Complex Resource}
\newacronym{PNF}{PNF}{Physical Network Function}
\newacronym{CP}{CP}{Connection Point}
\newacronym{VL}{VL}{Virtual Link}
\newacronym{SDC}{SDC}{Service Design and Creation}
\newacronym{ONAP}{ONAP}{Open Network Automation Platform}
\newacronym{VID}{VID}{Virtual Infrastructure Deployment}
\newacronym{VSP}{VSP}{Vendor Software Product}

\newacronym{WEF}{WEF}{Wireless Edge Factory}

\newacronym{DP}{DP}{Data Plane}
\newacronym{ECOMP}{ECOMP}{Enhanced Control Orchestration Management and Policy}

\newacronym{AAI}{AAI}{Active and Available Inventory}
\newacronym{SDNC}{SDNC}{Software Defined Network Controller}
\newacronym{SO}{SO}{Service Orchestrator}
\newacronym{APPC}{APPC}{Application Controller}
\newacronym{DCAE}{DCAE}{Data Collection Analytics and Events}
\newacronym{OOF}{OOF}{ONAP Optimization Framework}
\newacronym{OSS}{OSS}{Operation Support System}
\newacronym{BSS}{BSS}{Business Support System}
\newacronym{SOCKS}{SOCKS}{Secured Over Credential-based Keberos}
\newacronym{VVP}{VVP}{VNF Validation Program}
\newacronym{PDP}{PDP}{Policy Decision Point}
\newacronym{PEP}{PEP}{Policy Enforcement Point}
\newacronym{PCC}{PCC}{Policy Creation Component}
\newacronym{RU}{RU}{Remote Unit}

\newacronym{VLM}{VLM}{Vendor License Model}

\newacronym{I/Q}{I/Q}{in-phase / in-quadrature}
\newacronym{CUPS}{CUPS}{Control User Plane Separation}

\newglossarystyle{modsuper}{%
  \glossarystyle{super}%
}

\begin{document}

\title{Cloud-RAN functional split for an efficient fronthaul network}

\author{ ISBN: 978-1-7281-3130-6 \\ \IEEEauthorblockN{Veronica Quintuna Rodriguez\IEEEauthorrefmark{1}, Fabrice Guillemin\IEEEauthorrefmark{1},  Alexandre Ferrieux\IEEEauthorrefmark{1} and Laurent Thomas\IEEEauthorrefmark{2}}\\ 
\IEEEauthorblockA{\IEEEauthorrefmark{1}Orange Labs, 2 Avenue Pierre Marzin,  22300 Lannion, France} \IEEEauthorrefmark{2}Open Cells Project, 6 Chemin du Bois Brule, 91120 Palaiseau, France\\

}
\maketitle  

\begin{abstract}
The  evolution of telecommunication network towards cloud-native environments enables flexible centralization of the base band processing of radio signals. There is however a trade-off between the centralization benefits and the fronthaul cost for carrying the radio data between distributed antennas and data processing centers, which host the virtual RAN functions. In this paper, we present a specific  split solution for an efficient fronthaul, which enables reducing the consumed bandwidth while being compliant with advanced cooperative radio technologies (interference reduction and data rate improvements).  The proposed split has been implemented on the basis of Open Air Interface code and  shows important gains in the required fronthaul bandwidth as well as significant latency reduction in the processing of radio frames. 
\end{abstract}



\section{Introduction}

Virtualization techniques deeply modify the architecture of telecommunication networks, notably via \gls{NFV}~\cite{ETSI_NFVArch,NFV}. While network functions were so far bound to their hosting hardware, virtualization enables a full  decoupling between functions and hardware. While this change applies to  many network functions (mobile core network, firewalls, etc.), virtualization has  a particularly strong impact on  \gls{RAN} and especially \gls{C-RAN}.

The concept of \gls{C-RAN} was introduced a few years ago to take benefit of the centralization  of \gls{BBU} functions, for better radio and computing resource allocation, energy efficiency, cost-saving, etc~\cite{chinaMobile}.  Beyond resource efficiency \gls{C-RAN} shall bring agility, flexibility and time-to-market acceleration of future network features, which can  be implemented via software upgrades without implying hardware changes~\cite{cloudRAN_mec}.

Today, \gls{C-RAN} is a well established concept and has given rise to industrial implementations \cite{vRANwindWP2017,Intel_Altiostar}. An O-RAN alliance has been founded by operators to release open interfaces and to accelerate the delivery of products~\cite{ORANWG4-CUS}. In general, a \gls{C-RAN} is composed of three entities: \gls{RU}, \gls{DU} and \gls{CU}. An \gls{RU} hosts lower physical (LoPhy) functions while a DU executes higher physical (HiPhy) functions, MAC procedures and radio control tasks. A CU performs the higher \gls{RAN} functions belonging to \gls{RRC} and \gls{PDCP} protocols. The location of RUs is as close as possible to antennas, while DUs are in general located higher in the network in order to perform coordination of radio resource allocation and execute complex tasks (radio scheduling, channel coding) requesting intensive computing. A CU can be more centralized, say, at the PoP level, to manage, for instance, hand over so as to get rid of X2 interface between gNodeBs.

From a business perspective, \gls{C-RAN}  perfectly fits the emerging  needs of TowerCos~\cite{cloudRAN_mec}. The owners of towers equipped or not with data centers, can host RU functions while DUs and CUs can be located in data centers in the backhaul or even higher in the network. For a network operator, it becomes possible to install DU/CUs on data centers and to maintain them with classical software upgrade/update tools. Elementary RAN functions are then executed in dedicated RUs.

To reach this goal, two critical aspects have to be addressed in the design of \gls{C-RAN}: the fronthaul bandwidth and the distance between RUs and their controlling DUs (fronthaul size). The first point is deeply related to the functional split between HiPhy and LoPhy functions. A detailed description of different split options are given in the 3GPP specification~\cite{3GPP38_801}. Classical virtualized RAN (vRAN) was based on  Option 8 that leads to huge fronthaul bandwidth capacities  (about 10 Gbps per radio cell of 20~MHz). Meanwhile, ORAN alliance~\cite{ORANWG4-CUS}~\cite{oranWebSite} has more specifically considered Option 7.2. We shall consider this split as a reference in the following.

\begin{figure}[hbtp]
\centering  \includegraphics[scale=0.46, trim=65 10 20 20, clip]{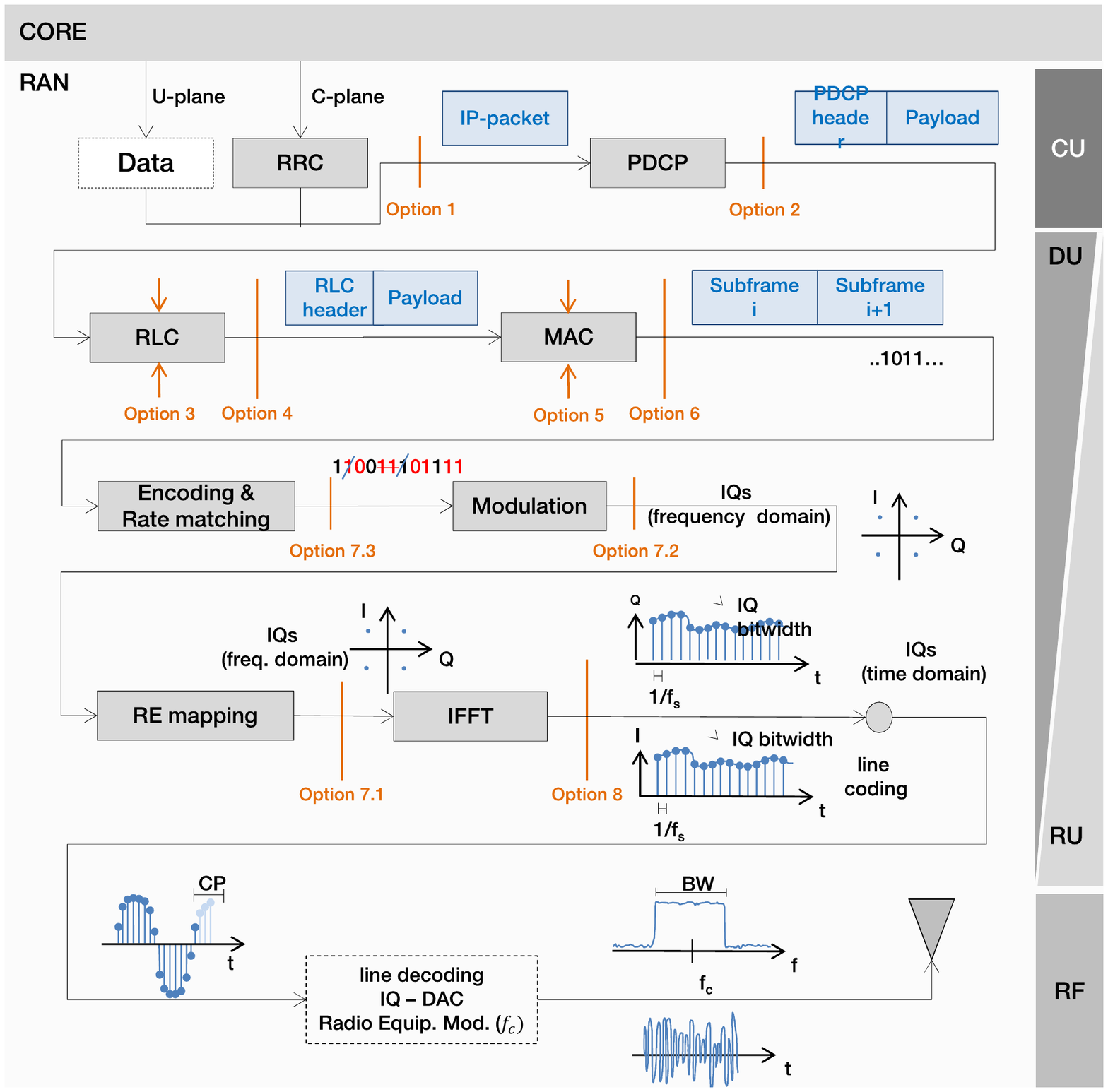}
\caption{C-RAN functional splits (downlink view).\label{fsplit}}
\end{figure}

On the other hand, fronthaul size is constrained by the standardized \gls{RTT} budget which enables specifying response times and retransmission periods. In LTE the \gls{HARQ} loop standardizes the round-trip latency to 3 milliseconds, then radio frames must to be processed within 1 and 2 milliseconds in the downlink and uplink directions, respectively, in order to meet the loop constraints. Thus, the distance between RUs and DUs directly depends on the computing platform and software optimization of RAN functions. In order to increase the fronthaul size, radio frames need to be processed as fast as possible. This calls for multi-core and multi-threading architectures. Remaining time after the base band processing is available for transmitting the radio information between the DU and RU at rate of 5 microseconds per kilometer (the light speed in optical fibers is taken equal to 200,000 km/s as an approximation).  

In this paper, we pay special attention to the 7.3 split and moreover we introduce the uplink direction which has so far not been considered in the state of the art. This is motivated by the need to limit  the fronthaul bandwidth. As a matter of fact, only hard bits are transmitted in the downlink; this is much more efficient than transmitting I/Q signals (e.g. splits 7.2) and  naturally scales to goodput. Similarly, in the uplink the split consists of transmitting soft bits  instead of radio symbols. The proposed split notably avoids compressing I/Q signals for the reduction of the required fronthaul capacity, in order to prevent additional signal degradation. Furthermore, by considering multi-threading techniques on a multi-core platform, it is possible to concentrate DUs several tens of kilometers from \glspl{RU}.

This paper is organized as follows: In Section~\ref{rationale}, we review the various  functional splits so far studied by standardization bodies as well as by Industry; this leads us to select Option 7.3. This split is further detailed  in Section~\ref{sec:fs73}. Testbed implementation guidelines and experimental results are presented in Section~\ref{sec:testbed}. Concluding remarks are finally presented in Section~\ref{sec:conclusion}.

\section{Rationale for the choice of a functional split}
\label{rationale}

\subsection{Functional split options}
Several  options for splitting RAN functions have been studied in the literature and notably by 3GPP~\cite{3GPP38_801}, which proposes eight splits going from Option 1 (higher-layer split) to Option 8 (lower-layer split); see Figure~\ref{fsplit}. A fully centralized architecture (Option 8) has the benefit of cost reduction due to a fewer number of sites hosting RAN functions but requires high fronthaul capacity to transmit radio signals on fiber. The fronthaul capacity problem resides not only in the constant bit rate used by the legacy fronthaul protocol, namely \gls{CPRI}~\cite{duan2016performance}, but also in the high redundancy present in the transmitted I/Q signals.

The assessment of the various splits has required huge research efforts; a survey of them is given in~\cite{larsen2018survey}. Particular attention has been paid to the intra-PHY functional splits due to the centralization advantages that they can achieve. Thus, most research efforts have turned to compression methods for reducing the required fronthaul capacity. Higher splits are less interesting since they do not exploit the centralization benefits and involve complex remote units.

The main features of the various splits are summarized in Table~\ref{tab:Splits}, which clearly shows the trade-off between RU complexity and required fronthaul bandwidth. It is quite clear that less air interface optimization is possible from DU coordination when keeping RAN functions in the RU.

\begin{table}[hbtp]
\caption{Main features of C-RAN splits}
\label{tab:Splits}
\begin{center}
\begin{tabular}{L{0.55cm}L{3.3cm}L{3.3cm}}
\toprule
{Split} & {Advantage} & {Drawback}  \\ 
\midrule
8 & Lowest RU complexity, highest centralization & Very high \gls{CBR} in the fronthaul scaling with antennas\\ 
7.1 & Low RU complexity & Fronthaul bandwidth scales with the number of antennas \\ 
7.2 & Fronthaul bandwidth scales with used spectrum instead of number of antennas & High \gls{CBR} in the fronthaul\\ 
7.3DL& Low and load dependent bitrate on the fronthaul while keeping high centralization & No symmetry, 7.3 uplink is not considered by 3GPP\\ 
6& Centralized scheduling& The close relation between FEC and MAC disappears \\ 
5& Real-time functions are in the DU & Complex interface\\ 
4& Low fronthaul data rate& The close relation between RLC and MAC disappears \\ 
3& Favors reliability& Latency sensitive in some cases\\ 
2& Enables mobility coordination & Low centralization\\ 
1& User plane separation & Complex RU\\ 
\bottomrule
\end{tabular}
\end{center}
\end{table}

It is worth noting that the complexity of the RU is intrinsically related to the way of implementing LoPhy functions. For instance, by using FPGA for coding and Fourier transforms,  the complexity of the RU clearly decreases as well as the associated energy consumption, at the expense of less agility with regard to software upgrades. Split 6 could thus be envisioned, but at the expense of losing the co-location of FEC and MAC (for real-time cooperative transmissions) and the joint processing of path diversity in the DU.

It turns out that 7.x family splits offer the best balance between RU complexity and inter-cell cooperation, though splits 7.1 and 7.2 can be greedy in terms bandwidth. We more thoroughly analyze 7.x splits in the next section.

\subsection{Analysis of 7.x splits}
\label{7xplits}

The assessment of intra-PHY options leads us to consider the fronthaul bandwidth as a critical \gls{KPI}. In fact, the various 7.x options enable beam-forming, inter-cell coordination, low RU complexity, etc. and are future-proof as they allow the  introduction of  new features via software upgrades.

The required transmission bandwidth of 7.x splits is mainly determined by physical features such as:
\begin{itemize}
    \item The cell bandwidth $BW_{cell}$ and the number of sub-carriers $N_{sc}$ (e.g., 1200 sub-carriers for a cell of 20~MHz).
    \item The modulation order $O_m$, number of bits per symbol (e.g. 4 bits for 16 QAM).
    \item The number of MIMO layers $N_{layers}$ (e.g., 2,4,8).
    \item The number of antenna ports\footnote{Antenna ports are logical entities and do not correspond to physical antennas. Various antenna port signals can be transmitted on a single physical antenna. Similarly a single port signal can be spread across various antennas.} $N_{ant}$. 
    \item The I/Q size $IQ_{bw}$, i.e, the required number of bits to code a constellation point (e.g., 32 bits for both in-phase and in-quadrature data). 
\end{itemize}

Recent studies aim at reducing the I/Q size by compression methods, which involves floating point compressed representations, $\mu$-law application, modulation compression schemes, and others (see for instance~\cite{guo2013lte}). The ORAN initiative has notably included in the fronthaul  ORAN-WG4.CUS.0-v0.2.0~\cite{ORANWG4-CUS} specification the possibility of transmitting I/Q data for both DL and UL (including user data and control channels) in compressed format. The specification is based on various compression and decompression techniques; however, selection algorithms and their performance evaluation are not yet available. 

The main drawback when compressing I/Q signals is that they contain sensitive radio information elements, which are particularly critical for the RAN intelligence in the goal of reducing the noise impacts of the radio channel. Losing precision on the I/Q symbols is then highly risky for the user payload data rate and QoS of the RAN network.

\subsubsection{7.1 split}
This split consists of transmitting I/Q symbols in the frequency domain. This option saves the overhead introduced by the frequency to time conversion \footnote{The oversampling factor is  given by  $F_{oversampling}={N_{FFT}}/{N_{sc}}=~1.71$, where $N_{FFT}$ is the number of FFT samples per symbol and $N_{sc}$ is the number of sub-carriers per cell.}. An experimental evaluation of low splits, namely Option 8 (referred to as IF5) and Option 7.1 (referred to as 4.5) is in~\cite{Experimental71}.
Roughly, the  fronthaul capacity $R_{7.1}$ (in bit/s, denoted for short  as bps) required by the 7.1 split  is mainly determined by the symbol size and the number for antenna ports as follows: 
\begin{equation}
\label{ec:s71}
    R_{7.1} = \frac{2*IQ_{bw}*N_{sc}*N_{ant}*N_{layers}}{T_s}.
\end{equation}
where $T_s$ is the symbol period given by the number of symbols carried in a time slot. For instance, when using normal cyclic prefix,  LTE transmits $7$ symbols per slot of $0.5$ milliseconds ($T_s=0.5/7=0.07$ milliseconds).

\subsubsection{ 7.2 split}
Like 7.1 split, 7.2 split transmits I/Q signals in the frequency domain. However, signals coming from multiple antenna ports are combined; as a consequence, the required fronthaul capacity $R_{7.2}$  (in bps) is divided by $N_{ant}$, so that
\begin{equation}
\label{ec:s72}
  R_{7.2}=\frac{2*IQ_{bw}*N_{sc}*N_{layers}}{T_s}
\end{equation}

\subsubsection{7.3 split}

By keeping the demodulation/modulation function near to antennas, the fronthaul with the 7.3 split carries bits instead I/Q symbols. The required fronthaul capacity $R_{7.2}$ (in bps) is then divided by $2*IQ_{bw}$.  In the downlink, the fronthaul capacity is determined by
\begin{equation}
\label{ec:s73DL}
    R_{7.3 DL} =\frac{N_{sc}*N_{layers}*O_m}{T_s}
\end{equation}

Table~\ref{tab:capacitySplits} shows the required fronthaul capacity for the 7.x splits for a cell bandwidth of 10~MHz and 20 ~MHz, 2x2~MIMO ($N_{layers}=2)$, 4 antenna ports ($N_{ant}=4$) and $16$~QAM modulation ($O_m=4$). Option 8 is indicated as a reference.

\begin{table}[hbtp]
\caption{Required Fronthaul Capacity}
\label{tab:capacitySplits}
\begin{center}
\begin{tabular}{L{1.6cm}R{2cm}R{2cm}}
\toprule
{Split} & $R_{x}$ [Mbps]   & $R_{x}$ [Mbps]  \\ 
 & (10 MHz) &  (20 MHz)\\ 
\midrule
Option 8  &$3677.2$& $7357.4$ \\ 
Option 7.1&$2150.4$ & $4300.8$ \\ 
Option 7.2&$537.6$ & $1075.2$ \\ 
Option 7.3&$67.2$ & $134.4$ \\ 

\bottomrule
\end{tabular}
\end{center}
\end{table}

As observed in Table~\ref{tab:capacitySplits}, when using 7.1 split, the fronthaul capacity is halved when compared with the initial \gls{CPRI}-based solution (Option 8).  This however still leads to a large required fronthaul capacity. The 7.3 split is certainly  the least bandwidth consuming.  Nevertheless, 3GPP considers 7.3 only in the downlink direction. We introduce in Section~\ref{sec:fs73} the corresponding 7.3 uplink split.

\section{7.3 split implementation}
\label{sec:fs73}

\subsection{Principles}

\begin{figure}[hbtp]
\centering
  \includegraphics[scale=0.46, trim=0 180 170 0, clip] {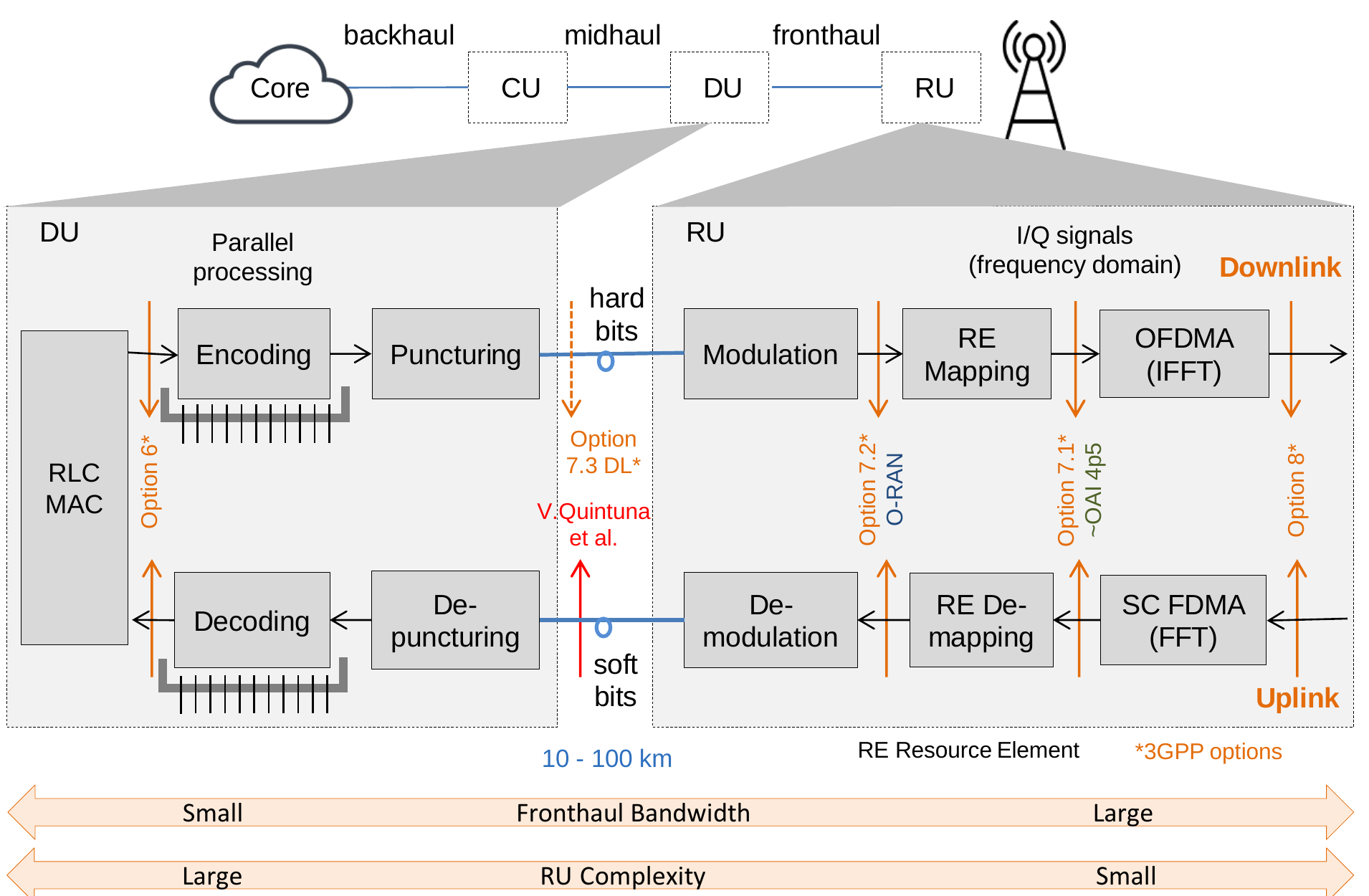}
    \caption{Implemented functional split architecture.}
    \label{fig:fs73}
\end{figure}

In this section, we describe the architecture associated with 7.3 split for both uplink and downlink directions. The proposed split centralizes the RAN upper layers including the channel  encoding/decoding function higher in the network and distributes near to antennas (RU) the lower PHY functions, i.e., modulation/demodulation and OFDM signal generation/reception~\cite{inriaMyPhD,icin2019}. See Figure~\ref{fig:fs73} for an illustration.

Hard bits are sent in the downlink while soft bits (coded as real numbers) are transmitted  in the uplink. Soft bits corresponds to \gls{LLR}, i.e., the log ratio of the probability that a particular bit is 1 to the probability that the same bit is 0 (logarithmic ratios are used for better precision). Soft bits need to be coded with a given precision referred to as bitwidth or size (e.g., 8 bits). Both hard and soft bits are encapsulated into UDP packets and transmitted on fiber links.  The required fronthaul bandwidth in the uplink direction is  given as by
\begin{equation}
\label{ec:s73UL}
    R_{7.3 UL} =\frac{N_{sc}*N_{layers}*O_m*S_{bw}}{T_s},
\end{equation}
where $S_{bw}$ is the soft bit size. 

\subsection{Benefits of 7.3 split}

7.3 split prevents from transmitting the whole cell bandwidth and combines the various radio signals coming from multiple antennas but first of all it transmits bits instead of I/Q signals. This yields significant reduction in the required fronthaul capacity. 

Beyond bandwidth, the fronthaul interface considered in the  7.3 split uses generic packet-based transport protocols instead of evolved vendor proprietary solutions, e.g., those based on \gls{CPRI}.

\subsection{Transport and Protocol Architecture}

We rely on  UDP as basic  transport mechanism between the RU and DU; we actually use the facilities of Linux UDP sockets. Sending \textit{sendSubFrame()} and receiving \textit{receiveSubFrame()} procedures are then used for both uplink and downlink directions.  While the transmission function sends UDP chunks as soon as possible, the reception mechanism aims at collecting all UDP chunks belonging to a given subframe before to determine: (i) the entire subframe has been successfully received, (ii) the timeout is expired before subframe completion, (iii) chunks are jumbled, i.e. a chunk from the next subframe is arrived. 

\begin{figure}[hbtp]
\centering
  \includegraphics[scale=0.46, trim=0 399 170 0, clip] {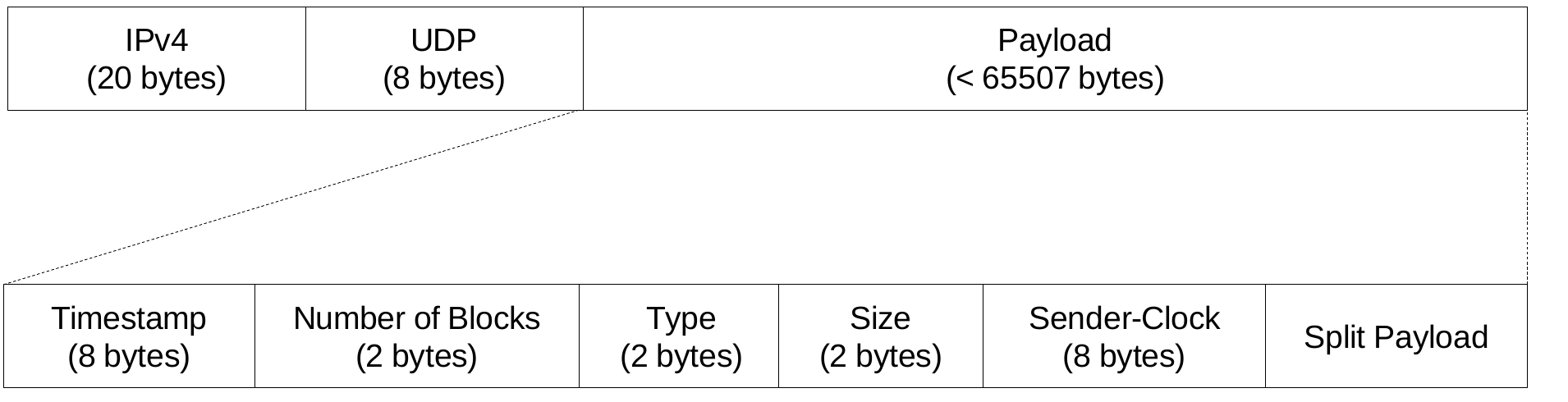}
    \caption{Split transport header.}
    \label{fig:header}
\end{figure}

The transport header includes (see Figure~\ref{fig:header}):
\begin{itemize}
    \item \textit{Timestamp}, subframe identifier (8 bytes).
    \item \textit{Number of blocks} belonging to the subframe (2 bytes).
    \item \textit{Type} which defines the content format (2 bytes).
    \item \textit{Size}\footnote{Theoretically, it must be less than $65,508$ bytes (from $2^{16}-8-20$ bytes  for UDP limit, UDP header and IP header, respectively).}, number of transmitted bytes including the split header (2 bytes). 
    \item \textit{Sender-Clock} (8 bytes).
\end{itemize}

Regarding LTE control information, the DU notably transmits resource block position, power and \gls{MCS} data with each data plane packet, which are used by the RU in low PHY functions. The RU in turn reports the \gls{CQI} information obtained from the signal processing. 

The DU sends signaling channels (e.g., DCI channel containing UE control information as HARQ feedback, scheduling information, frame organization, UE power control, UE paging requests, etc) to the RU as semantic data (number of DCI, position in the subframe, PHICH, PDCCH, ...).

\subsection{Worst-case Analysis: I/Qs versus bits}

For evaluating the gain that can be obtained when transmitting bits instead of I/Qs, we focus now on a worst-case analysis. Since the fronthaul bandwidth required by 7.3 split varies with the traffic load, we consider as a worst-case the peak data rate capacity of a cell of 100~MHz. Furthermore, the bandwidth saving achieved with  7.3 split strongly depends on the modulation order, we then consider the highest modulation schemes in Table~\ref{tab:bandwidthsORAN}~\cite{3GPP38_801,ORANWG4-CUS}.  In addition, we use  8 MIMO layers,  I/Q size of 2x16 bits, 32 antenna ports, and soft bit size of 5 bits.

\begin{table}[hbtp]
\caption{Required fronthaul bandwidth [Gbps], worst-case}
\label{tab:bandwidthsORAN}
\begin{center}
\begin{tabular}{C{0.4cm}C{1.6cm}C{1.6cm}C{1.6cm}}
\toprule
&Modulation&Option 7.3 & {Option 7.2}  \\ 
\midrule
DL& 256 QAM &4.1&22.2\\ 
UL& 64 QAM &20.25 &21.6\\ 
\bottomrule
\end{tabular}
\end{center}
\end{table}

\begin{figure}[hbtp]
\centering
  \includegraphics[scale=0.65, trim=0 270 320 0, clip] {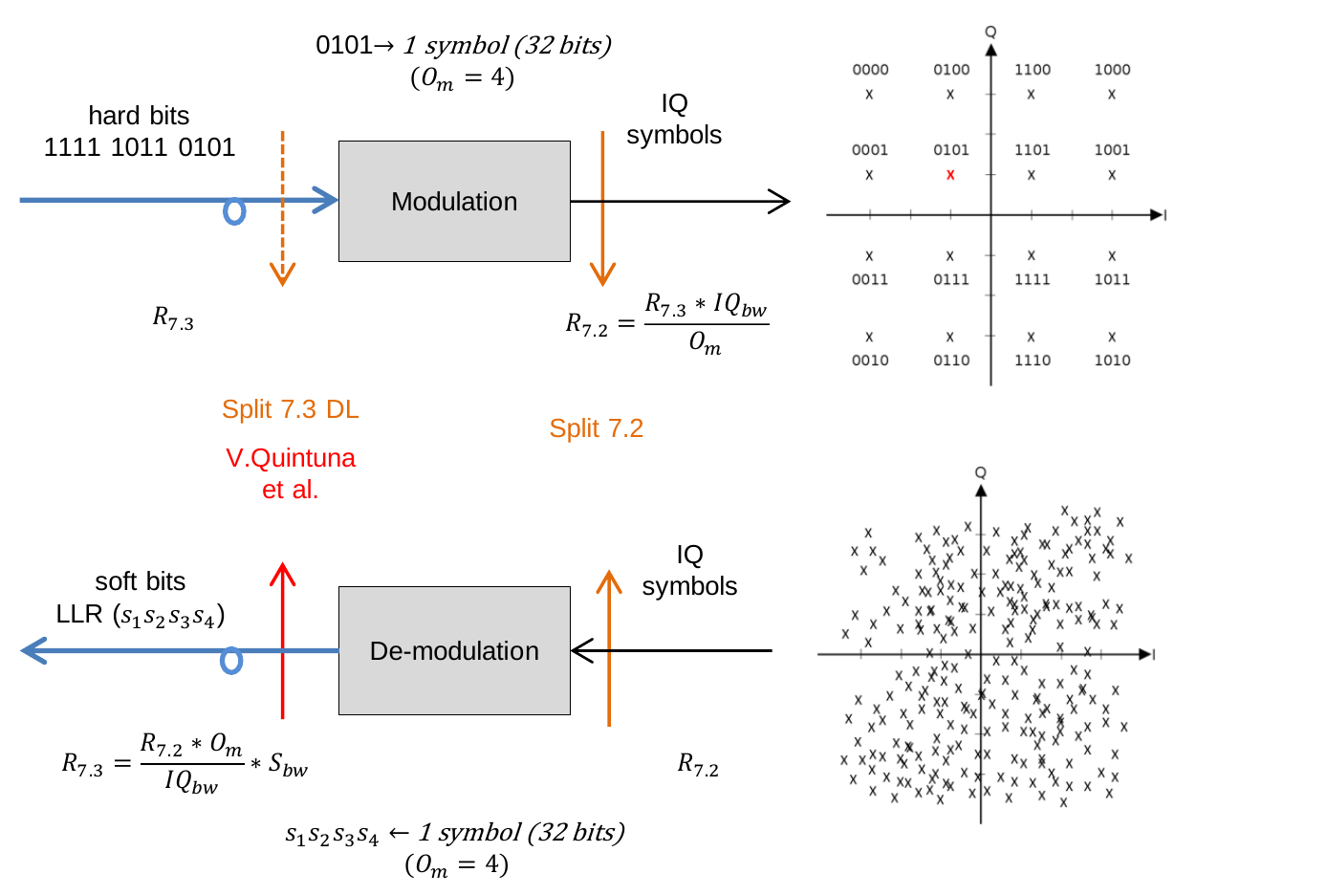}
    \caption{Fronthaul efficiency (7.2 vs. 7.3).}
    \label{fig:gains}
\end{figure}

Option 7.3 evidences important downlink fronthaul reduction (even in a worst-case scenario). The gain in the uplink can be quickly improved when varying the modulation type. As shown in Figure~\ref{fig:gains}, the trade-off between bits and I/Qs transmission is notably given by the modulation order.

Continuing with a worst-case approach (peak data rate), the overhead factor introduced when sending I/Q signals instead bits for various modulation orders is shown in Table~\ref{tab:gain}. It turns out that 7.3 split is up to $16$ times better than 7.2 split, when using a QPSK modulation in the downlink direction. In the uplink the required fronthaul bandwidth is up to $4$ times better when using 7.3 (4 bits size) and QPSK modulation. In practice, the most commonly used modulation orders in the uplink are 16 QAM and QPSK. For instance, by using statistics from an operating mobile network,  a  4G cell registers $36.35 \%$ and $28.43\%$ usage of QPSK and 16 QAM modulation schemes, respectively. Only $2.74\%$ of traffic is modulated in 256 QAM.

\begin{table}[hbtp]
\caption{Fronthaul efficiency (7.2 vs. 7.3)}
\label{tab:gain}
\begin{center}
\begin{tabular}{C{0.7cm}C{1.3cm}C{1.6cm}C{1.6cm}C{1.2cm}}
\toprule
$O_m$&Mod. Scheme&$7.2/7.3$ Uplink ($S_{bw}=8)$ &{$7.2/7.3$ Uplink ($S_{bw}=4)$} & {$7.2/7.3$ Downlink} \\ 
\midrule
2& QPSK &2&4&16\\ 
4& 16 QAM&1&2&8\\ 
6& 64 QAM&0.7&1.3&5.3\\ 
8& 256 QAM&0.5&1&4\\ 
\bottomrule
\end{tabular}
\end{center}
\end{table}

Beyond the modulation order, the uplink savings can take advantage of the size used for coding soft bits. Unlike I/Qs, soft bits size can be considerably reduced while avoiding signal degradation.  

In the next section, we report experimental results from a testbed implementing the 7.3 split. The objective is to compare the theoretical estimations derived in the previous  sections against experimental results.

\section{Testbed experiments}
\label{sec:testbed}

\subsection{Settings}

In this section, we evaluate the performance of the proposed functional split (bidirectional 7.3 split) via a testbed implementation. The platform is notably composed of  a distributed entity (placed near to antennas) and a centralized one (located in the backhaul network), which  host the RU and the CU-DU, respectively. In this proof of concept, the RU and DU are linked by $10$ kilometer long  optical fiber. The CU is co-localized with DU. 

\begin{figure}[hbtp]
\centering
  \includegraphics[scale=0.46, trim=0 170 170 0, clip] {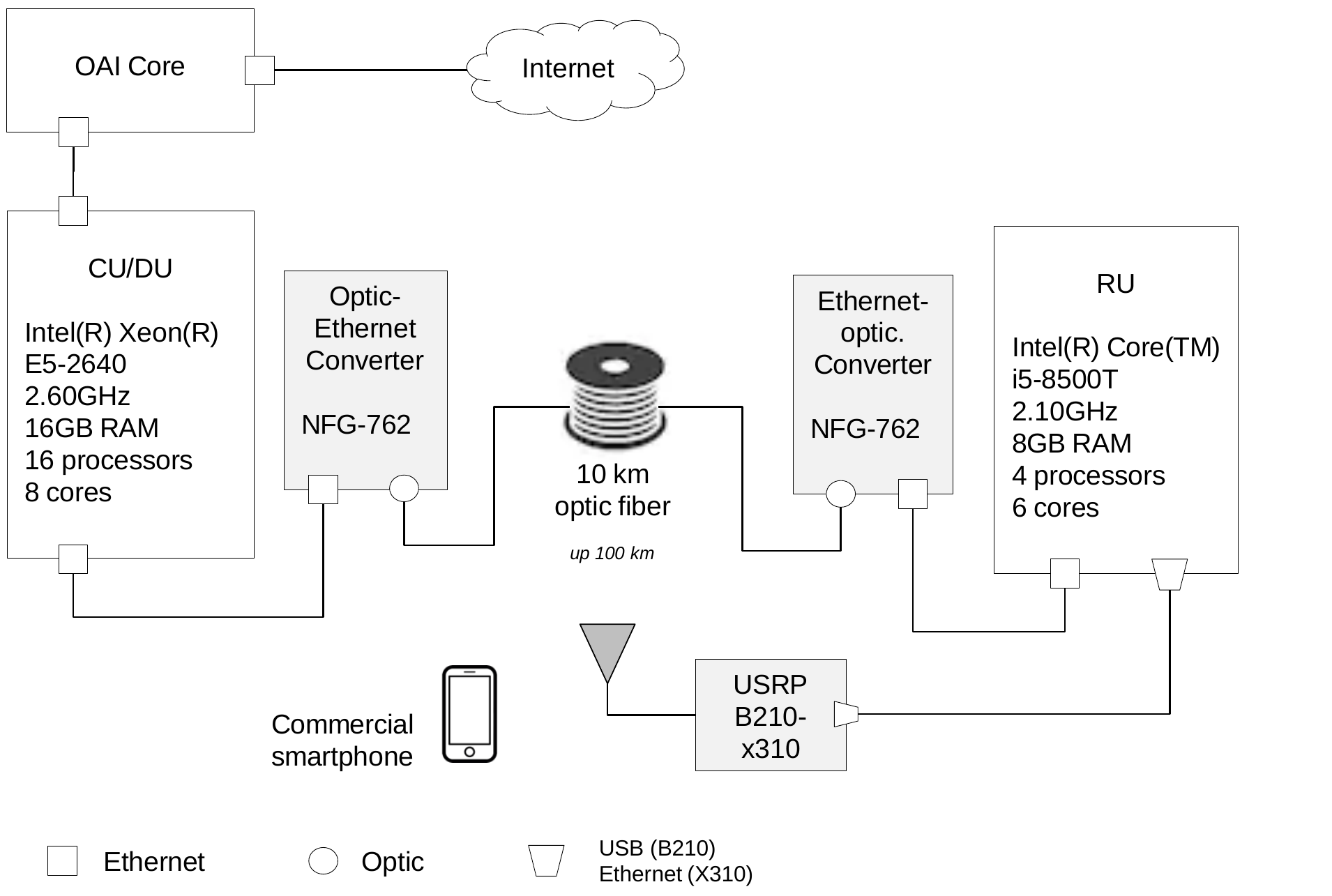}
  \caption{Testbed architecture.}
    \label{fig:testbed}
\end{figure}

The RAN functions for the three RAN units, RU, DU, CU are based on \gls{OAI}~\cite{oaiWebSite} code and run on general purpose computers. The radio cell is configured in \gls{FDD} mode and has  a 10~MHz bandwidth. The \gls{RF} module is an USRP X310 card (tests have been also carried out when using an USRP B210). The testbed architecture and main features are presented in Figure~\ref{fig:testbed}.

The main challenges when implementing intra-PHY splits are not only on the fronthaul interface but on the high performance computing required to meet the RAN real-time constraints. To reduce the processing time in the DU and thus to increase the distance between the RU and the DU, we  implement in the DU the multi-threading model described in~\cite{inriaMyPhD}, which by means of parallel processing and adapted scheduling principles~\cite{jsac} achieve an important reduction in the runtime of the channel encoding and decoding (most resource consuming) functions.  The thread-pool implementation is presented in~\cite{icin2019}. The achieved RAN acceleration justifies the co-location of the CU/DU in the backhaul network (fronthaul size up 100 kilometers).

\subsection{Fronthaul evaluation}

In order to evaluate the required fronthaul capacity in the DU-RU interface, we vary the useful data rate (user traffic) from $~0$ (when using \textit{ping}, often 32 or 56 byte long) up to the peak data rate for a cell at 10~MHz. We concretely make use of a traffic generator, which sends UDP packets at a configurable rate to guarantee the chosen throughput regardless of random packet loss. These UDP packets are handled by the complete RAN protocol stack (PDCP, RLC, etc.) in both downlink (by the DU) and uplink (by the UE) directions. 

\begin{figure}
\centering
  \includegraphics[scale=0.65, trim=100 260 100 280, clip]   {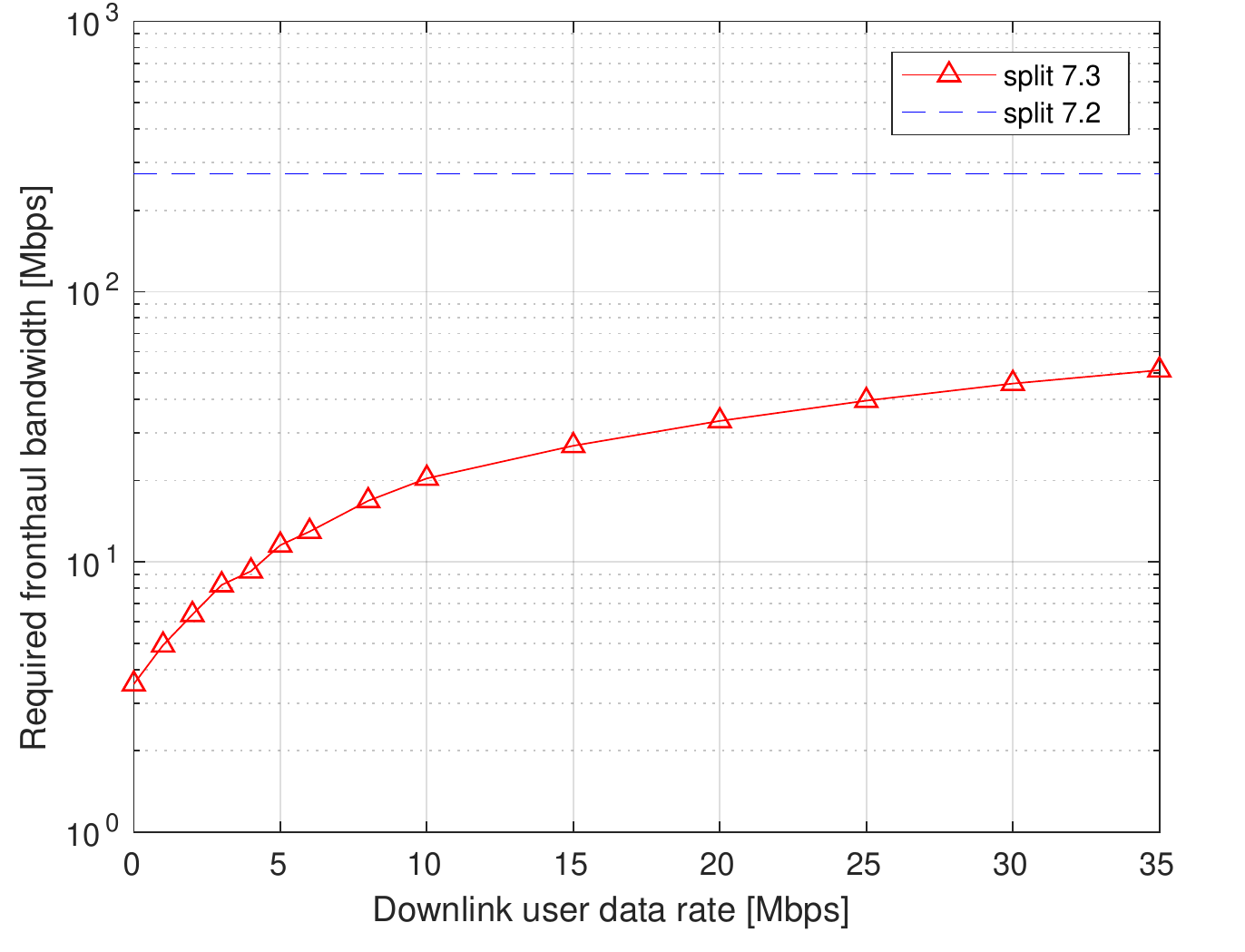}
  \caption{Required fronthaul capacity (downlink).}
   \label{fig:downlink}
\end{figure}

\begin{figure}
\centering
 \includegraphics[scale=0.65, trim=100 260 100 280, clip] {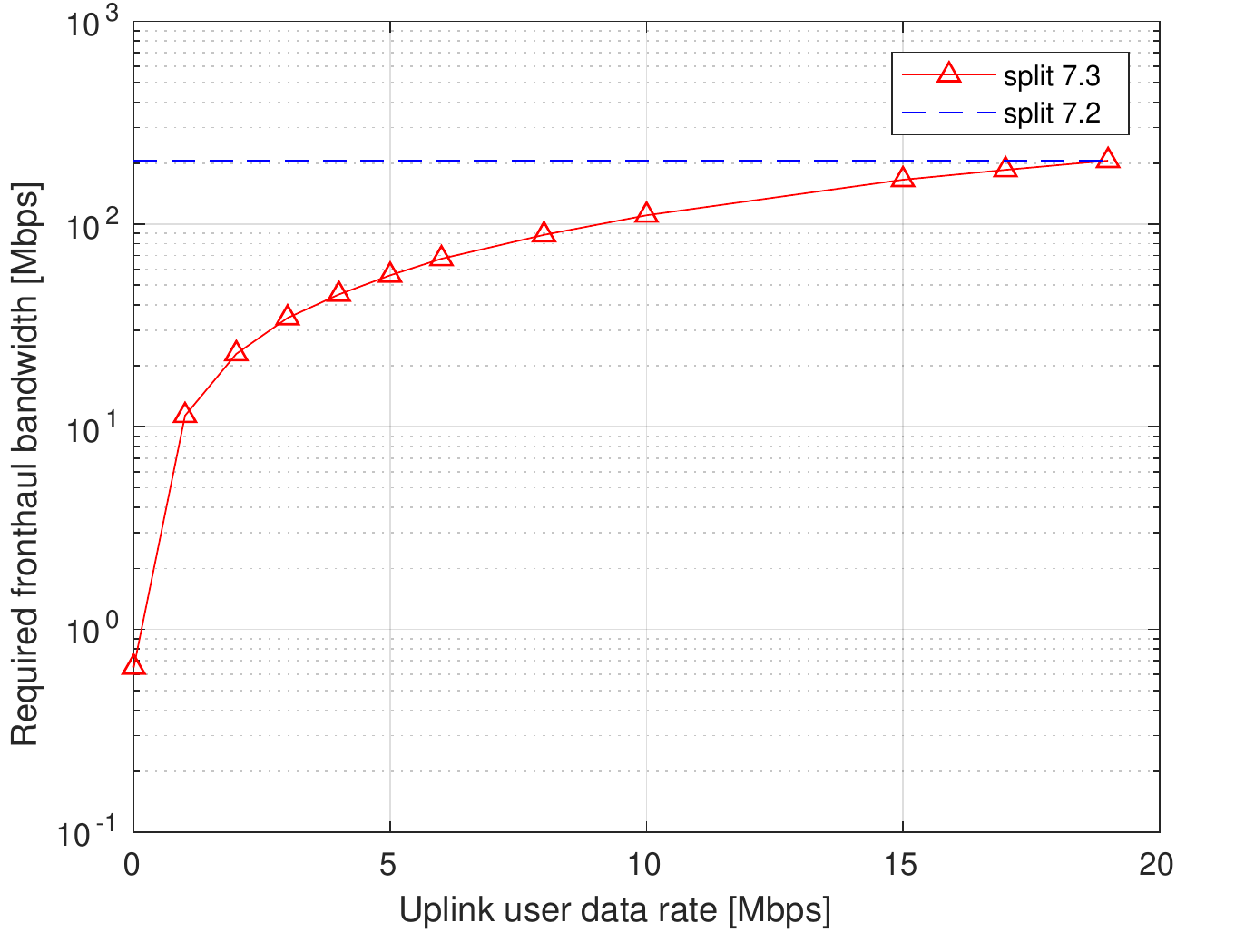}
   \caption{Required Fronthaul capacity (uplink).}
    \label{fig:uplink}
\end{figure}

The measured bandwidth in the fronthaul interface for the downlink (hard bits) and uplink (soft bits of 8 bits)  is  shown in Figures~\ref{fig:downlink} and~\ref{fig:uplink}, respectively.  The required fronthaul capacity is represented as a function of the traffic data rate. 

Performance results show important gains in the required fronthaul bandwidth with respect to other intra-PHY solutions (namely,  7.2 and 7.1 options of 3GPP~\cite{3GPP38_801}). In addition, it is observed that for both the uplink and downlink directions the required bandwidth varies with the useful data rate. Thus,  fronthaul dimensioning can take advantage of variable bit rates by implementing statistical multiplexing and resource sharing mechanisms for important capital cost reduction.

Additionally, the split uses a packet based system for transmitting hard bits in the downlink direction and soft-bits in the uplink. Replacing the \gls{CPRI} transmission over raw synchronous optical links, by packets over UDP/IP enables meshed connectivity between RAN components and allows effective adaptable and programmable networks~\cite{fujitsuCloudRANWP}.

\section{Conclusion}
\label{sec:conclusion}

We have addressed in this paper the analysis and implementation of the 7.3 intra-PHY functional split in a C-RAN architecture and compared it with those of 7.x family, namely 7.1 and 7.2 splits, this latter being adopted by the ORAN alliance. We have particularly introduced the uplink 7.3 split in order to keep the centralization benefits while reducing the required fronthaul capacity (RU-DU interface). Achieved symmetry simplifies the 7.3 fronthaul interface and enables us to use UDP as basic transport mechanism. It opens the door to meshed connectivity and effective adaptable networks. 

It turns out that the 7.3 split outperforms the other ones in terms of required fronthaul bandwidth and achievable distance between RU and DU, notably when considering suitable multi-threading of (de)coding functions.

The proposed split has been implemented on the basis of OAI open source code and tested in a proof of concept. The experimental results are in concordance with the theoretical estimations of the bit rates needed on the transmission link between the RU and DU. The next step is to integrate this split in the global RAN management architecture specified by ORAN, notably the \gls{RIC}.

\bibliographystyle{IEEEtran}
\bibliography{biblo}
\flushend
\pagebreak

\end{document}